\documentclass[a4paper, 12pt, notoc]{article}
\usepackage{epsfig,amssymb,euscript,xspace, color}
\usepackage{amsmath,jheppub}
\def\bea{\begin{eqnarray}}
\def\eea{\end{eqnarray}}
\def\be{\begin{equation}}
\def\ee{\end{equation}}



\DeclareMathOperator\sech{sech}
\DeclareMathOperator\csch{csch}
\newcommand\AdS{$AdS_3$\xspace}
\newcommand\re{\mathbb{R}}
\newcommand\sacomment[1]{\textcolor{blue}{[\textit{SA: #1}]}}
\notoc

\begin{document}
\title{
An $sl(2,{\mathbb R})$ current algebra from $AdS_3$ gravity}
\author{Steven G. Avery, Rohan R. Poojary and  Nemani V. Suryanarayana}
\affiliation{Institute of Mathematical Sciences \\ CIT Campus, Taramani \\ Chennai 600113, India}
\emailAdd{savery,ronp,nemani(at)imsc.res.in}
%\date{}

\abstract{ We provide a set of chiral boundary conditions for three-dimensional gravity that allow
  for asymptotic symmetries identical to those of two-dimensional induced gravity in light-cone
  gauge considered by Polyakov. These are the most general boundary conditions consistent with the
  boundary terms introduced by Comp\`{e}re, Song and Strominger recently. We show that the asymptotic
  symmetry algebra of our boundary conditions is an $sl(2, \mathbb{R})$ current algebra with level
  given by $c/6$. The fully non-linear solution in Fefferman--Graham coordinates is also provided
  along with its charges. }

\maketitle

%\tableofcontents

\newpage
\section{Introduction}

In his seminal 1987 paper \cite{Polyakov:1987zb}, Polyakov provides a solution to the
two-dimensional induced gravity theory \cite{Polyakov:1981rd},
\begin{equation}
\label{polyeq1}
S = \frac{c}{96 \pi} \int d^2x \sqrt{-g} \, R \frac{1}{\nabla^2} R,
\end{equation}
by working in a light-cone gauge. The gauge choice puts  the metric into the form
\begin{equation}
\label{polyeq2}
ds^2 = - dx^+ dx^- + F(x^+, x^-) (dx^+)^2.
\end{equation}
Polyakov shows that the quantum theory for the dynamical field $F(x^+, x^-)$ admits an $sl(2,
\mathbb{R})$ current algebra symmetry with level $k= c/6$. In this note, we present the
three-dimensional bulk theory that is dual to this two-dimensional theory.

\section{Chiral boundary conditions in $AdS_3$ gravity}

The action of three-dimensional gravity with negative cosmological
constant~\cite{Balasubramanian:1999re} is given by
\bea
S = - \frac{1}{16 \pi G} \int d^3x \, \sqrt{-g} \left( R - \frac{2}{l^2} \right) - \frac{1}{8 \pi G} \int_{\partial {\cal M}} d^2 x \, \sqrt{-\gamma} \, \Theta + \frac{1}{8 \pi G} S_\text{ct} (\gamma_{\mu\nu}),
\eea
where $\gamma_{\mu\nu}$ is the induced metric and $\Theta$ is trace of the extrinsic curvature of the boundary. Varying the action yields
\bea
\delta S = \int_{\partial {\cal M}} d^2x \sqrt{-\gamma} \, \frac{1}{2} T^{\mu\nu} \delta \gamma_{\mu\nu} \, ,
\eea
where 
\bea
T^{\mu\nu} = \frac{1}{8\pi G} \left[ \Theta^{\mu\nu} - \Theta \gamma^{\mu\nu} + \frac{2}{\sqrt{-\gamma}}\frac{\delta S_{ct}}{\delta \gamma_{\mu\nu}} \right].
\eea
The variational principle is made well-defined  by imposing $\delta \gamma_{\mu\nu} = 0$ (Dirichlet) or
$T^{\mu\nu} = 0$ (Neumann) at the boundary (see~\cite{Compere:2008us} for a recent discussion).

Recently Comp\`{e}re, Song and Strominger (CSS)~\cite{Compere:2013aya, Compere:2013bya} and
Troessaert~\cite{Troessaert:2013fma} proposed new sets of boundary conditions for three-dimensional
gravity, which differ from the well-known Dirichlet-type Brown--Henneaux boundary
conditions~\cite{Brown:1986nw}.\footnote{In fact, the boundary conditions of
  \cite{Troessaert:2013fma} subsume those of~\cite{Brown:1986nw}.}  Before delving into specifics,
let us discuss the general strategy employed by~\cite{Compere:2013bya}. One begins by adding a term
of the type
\bea
\label{css-term}
S' = -\frac{1}{8 \pi G} \int_{\partial {\cal M}} d^2x \, \sqrt{-\gamma} \, \frac{1}{2} {\cal T}^{\mu\nu} \gamma_{\mu\nu}
\eea
for a fixed ($\gamma_{\mu\nu}$-independent) symmetric boundary tensor ${\cal T}^{\mu\nu}$. The
variation of this term is
\bea
\delta S' = - \frac{1}{8 \pi G} \int_{\partial {\cal M}} d^2x \, \sqrt{-\gamma} \tilde{\cal T}^{\mu\nu} \delta \gamma_{\mu\nu},
\eea
where $\tilde {\cal T}^{\mu\nu} =  {\cal T}^{\mu\nu} - \frac{1}{2} ({\cal T}^{\alpha\beta} \gamma_{\alpha\beta}) \,  \gamma^{\mu\nu} $. The variation of the total action then gives 
\bea
\label{totvar}
\delta S + \delta S' = \frac{1}{8 \pi G} \int_{\partial {\cal M}} d^2x \, \sqrt{-\gamma} (T^{\mu\nu} - \tilde{\cal T}^{\mu\nu}) \delta \gamma_{\mu\nu}.
\eea
Now the boundary conditions consistent with the variational principle depend on $\tilde {\cal
  T}^{\mu\nu}$. Generically, this leads to ``mixed'' type boundary conditions.  If
for a given class of boundary conditions some particular component of $T^{\alpha\beta}-\tilde{\cal
  T}^{\alpha\beta}$ vanishes sufficiently fast in the boundary limit such that its contribution to the integrand in \eqref{totvar} vanishes, then the corresponding component of
$\gamma_{\alpha\beta}$ can be allowed to fluctuate. Since we want the
 boundary metric to match~\eqref{polyeq2}, we would like Neumann boundary conditions for
$\gamma_{++}$. Therefore we choose $\cal T^{\mu\nu}$ such that the leading term of $T^{++}$ equals $\tilde {\cal T}^{++}$ in the boundary limit.

This condition has been imposed in \cite{Compere:2013bya}, with the addition of an extra boundary term \eqref{css-term} with\footnote{The induced metric $\gamma_{\mu\nu}$ differs from $g_{\mu\nu}^{(0)}$ of \cite{Compere:2013bya} by a factor of $r^2$. }
\begin{equation}\label{eq:css-T}
\mathcal{T}^{\mu\nu} = -\frac{1}{2r^4} N^2 l \delta^\mu_+\delta^\nu_+,
\end{equation}
and the following boundary conditions are imposed on the metric:
\begin{equation}\label{css} 
\begin{gathered}
g_{rr} = \frac{l^2}{r^2} + {\cal O}(r^{-4}), ~~ g_{r \pm} = {\cal O} (r^{-3}),\\ 
g_{+-} = - \frac{r^2}{2} + {\cal O}(r^0) ,~~g_{++} = r^2 f(x^+) + {\cal O}(r^0), ~~ g_{--} = - \frac{l^2}{4} N^2 + {\cal O}(r^{-1}),
\end{gathered}
\end{equation}
where $f(x^+)$ is a dynamical field and $N^2$ is fixed constant.\footnote{To relate to the notation
  in~\cite{Compere:2013bya}, set $N^2 = -\frac{16 G\Delta}{l}$ and $f(x^+) = l^2 \partial_+
  \bar{P}(x^+)$.} These boundary conditions give rise to an asymptotic symmetry algebra: a chiral
$U(1)$ current algebra with level determined by $N$. These also ensure that $T_{--}$
is held fixed in the variational problem, whereas $g_{++}$ is allowed to fluctuate as long as its
boundary value is independent of $x^-$.

In what follows, we show that~\eqref{css} are not the most general boundary conditions
consistent with the variational principle and the extra boundary term given by~\eqref{eq:css-T}. For this, we introduce a weaker set of consistent boundary conditions that enhance the asymptotic symmetry algebra to an
$sl(2, {\mathbb R})$ current algebra whose level is independent of $N$.

\subsection{New boundary conditions}
In the new boundary conditions, the class of allowed boundary metrics coincides with that
of~\eqref{polyeq2}. Since we want to allow $\gamma_{++}$ to fluctuate, we keep $T_{--}$
fixed in our asymptotically locally $AdS_3$ metrics. Therefore, we propose the following boundary conditions:
\begin{equation}\label{aps}
\begin{aligned}
g_{rr} &= \frac{l^2}{r^2} + {\cal O}(\tfrac{1}{r^4}), ~~ g_{r+} = {\cal O}(\tfrac{1}{r}), ~~ g_{r-} = {\cal O}(\tfrac{1}{r^3}), \\
g_{+-} &= - \frac{r^2}{2} + {\cal O}(r^0), ~~ g_{--} = - \frac{l^2 N^2}{4} + {\cal O}(\tfrac{1}{r}), \\
g_{++} &= r^2 F(x^+, x^-) + {\cal O} (r^0) ,
\end{aligned}
\end{equation}
where, as above, we take $F(x^+, x^-)$ to be a dynamical field and $N$ fixed.  The crucial
difference between these boundary conditions and those in~\eqref{css} is the different fall-off
condition for $g_{r+}$ which allows for the boundary component of $g_{++}$ to depend on $x^-$ as well. One must, of course, check the consistency of these conditions with the
equations of motion. This involves constructing the non-linear solution in an expansion in inverse
powers of $r$. Working to the first non-trivial order, one finds the following condition on $F(x^+,
x^-)$:
\be
N^2 \, \partial_- F(x^+, x^-) + \partial_-^3 F(x^+, x^-) = 0,
\ee
which forces $F(x^+, x^-)$ to take the form
\be
F(x^+, x^-) = f(x^+) + g(x^+) e^{i N x^-} + \bar g(x^+) e^{-i N x^-}
\ee
where $f(x^+)$ is a real function and $\bar g(x^+)$ is the complex conjugate of $g(x^+)$. 

Let us note that this is directly analogous to the form of $F(x^+, x^-)$ derived
in~\cite{Polyakov:1987zb}. Throughout our discussion we think of $\phi = \frac{x^+ - x^-}{2}$ as
$2\pi$-periodic (and $\tau = \tfrac{x^++x^-}{2}$ as the time coordinate), and therefore we restrict our consideration to $N\in\mathbb{Z}$. Similarly, we
impose periodic boundary conditions on $f(x^+)$ and $g(x^+)$.  If one takes the spatial part of the
boundary to be $\re$ instead of $S^1$, there are no such restrictions and one may even consider $N^2
< 0$ like in~\cite{Compere:2013bya}.

\subsection{The non-linear solution} 

One can write a general non-linear solution of $AdS_3$ gravity in Fefferman--Graham
coordinates  \cite{Skenderis:1999nb} as:
\begin{equation}
\label{nlsol1}
ds^2 = \frac{dr^2}{r^2} + r^2 \left[ g^{(0)}_{ab} + \frac{l^2}{r^2} \, g^{(2)}_{ab} + \frac{l^4}{r^4} g^{(4)}_{ab} \right] dx^a dx^b.
\end{equation}
The full non-linear solution with our boundary conditions is obtained when
\begin{equation}
\label{nlsol2}
\begin{aligned}
g^{(0)} _{++ } &= f(x^+) + g(x^+) \, e^{i N x^-} + \bar g (x^+) \, e^{-i N x^-}, ~~ g^{(0)}_{+-} = -\frac{1}{2}, ~~ g^{(0)}_{--} = 0, \\
g^{(2)}_{++} &= \kappa (x^+) + \frac{1}{2} N^2 \left[ g^2(x^+) \, e^{2i N x^-} 
        + \bar g^2(x^+) \,e^{-2i N x^-} \right] \\
        & ~~~~~~~~~~\,~+ \frac{i}{2} N \left[g'(x^+) e^{i N x^-} - \bar g'(x^+) e^{-i N x^-}\right] ,\\
g^{(2)}_{+-} &= \frac{1}{4} N^2 \left[ f(x^+) - g(x^+) \, e^{i N x^-} - \bar g (x^+) \, e^{-iN x^-}\right], 
      ~~ g^{(2)}_{--} = -\frac{1}{4} N^2,\\
g^{(4)}_{ab} &= \frac{1}{4} g^{(2)}_{ac} g_{(0)}^{cd} g^{(2)}_{db} \, ,
\end{aligned}\end{equation}
where in the last line $g_{(0)}^{cd}$ is $g^{(0)}_{cd}$ inverse.  As above, demanding that the
solution respects the periodicity of $\phi$-direction requires $N$ to be an integer and the functions
$f(x^+)$, $g(x^+)$ and $\kappa(x^+)$ to be periodic. This solution reduces to the one given in
\cite{Compere:2013bya} when $g(x^+) = \bar g(x^+) = 0$.
 
As mentioned in the previous subsection one can take $N$ to be purely imaginary when the boundary spatial coordinate is not periodic. In this case too the non-linear solution \eqref{nlsol1} continues to be a valid solution with $g(x^+)$ and $\bar g(x^+)$ treated as two real and independent functions. However, we will not consider this case further here.

\section{Charges, algebra and central charges}

It is easy to see that vectors of the form
\begin{equation}\begin{aligned}
\xi^r  &= -\frac{1}{2}\left[B'(x^+) +iN A(x^+) e^{iN x^-}
                       -iN \bar{A}(x^+)e^{iN x^-} \right]r  + \mathcal{O}(r^0)\\
\xi^+ &= B(x^+) - \frac{l^2N^2}{2r^2}\left[A(x^+)e^{iN x^-} 
                                     + \bar{A}(x^+) e^{-iN x^-}\right] + \mathcal{O}(\tfrac{1}{r^3})\\
\xi^-  &= A_0(x^+) + A(x^+)e^{iN x^-} + \bar{A}(x^+)e^{-iN x^-} + \mathcal{O}(\tfrac{1}{r})
\end{aligned}\end{equation}
%
%  \bea
%  V &=& - \frac{1}{2} B'_0(x^+) \, r\partial_r + B_0 (x^+) \partial_+ + \cdots \cr
%  && + A_0 (x^+) \partial_- + \cdots \cr
%  && + e^{i N x^-} A (x^+) \left[ \partial_- - \frac{l^2 N^2}{2r^2} \partial_+ - \frac{i}{2} N r\partial_r \right] + \cdots \cr
%  && + e^{-i N x^-} \bar{A} (x^+) \left[ \partial_- - \frac{l^2 N^2}{2r^2} \partial_+ + \frac{i}{2} N r\partial_r \right] + \cdots
%  \eea
%
satisfy the criteria of~\cite{Barnich:2001jy}, which allow us to construct corresponding asymptotic
charges. If, on the other hand, one demands that the asymptotic symmetry generators $\xi$ leave the
space of boundary conditions invariant, one finds the same vectors but with the first  subleading terms
appearing at one higher order for each component. For either set of vectors, the Lie bracket
algebra closes to the same order as one has defined the vectors.

Here, $B(x^+)$ and $A_0(x^+)$ are real and $A(x^+)$ is complex; therefore, there are four real,
periodic functions of $x^+$ that specify this asymptotic vector.
 %
%\bea
%&& f(x^+) \rightarrow -2 i N  g(x^+) \bar{A}(x^+)+2 i N 
%   A(x^+)
%   \bar{g}(x^+)-A_0'(x^+)+f(x^+)
%   B_0'(x^+)+ B_0 (x^+) f'(x^+) \cr
% &&  g(x^+) \rightarrow - \bar A'(x^+)-i N  \bar A(x^+) f(x^+)+ i N 
%   A_0 (x^+) \bar g(x^+)+ \bar g(x^+)
%   B_0'(x^+)+B_0(x^+) \bar g'(x^+) \cr
%    &&  \bar g(x^+) \rightarrow -A'(x^+)+i N  A(x^+) f(x^+)-i N 
%   A_0 (x^+) g(x^+)+g(x^+)
%   B_0'(x^+)+B_0(x^+) g'(x^+)
%\eea
%
We take the following basis for the modes of the vector fields:
\begin{equation}\begin{aligned}
L_n &= i e^{i \, n \, x^+}  [ \partial_+ - \frac{i}{2} n \, r\partial_r ] + \cdots \\
T^{(0)}_n &= \tfrac{i}{N} e^{i \, n \, x^+} \partial_- + \cdots \\
T^{(+)}_n &= \tfrac{i}{N} e^{i (n \, x^+ + N \, x^-)} [ \partial_- - \frac{i}{2} N \, r\partial_r  -\frac{N^2}{2r^2}\partial_+] + \cdots \\
T^{(-)}_n &= \tfrac{i}{N} e^{i (n \, x^+ - N \, x^-)} [ \partial_- + \frac{i}{2} N \, r\partial_r -\frac{N^2}{2r^2}\partial_+] + \cdots ,
\end{aligned}\end{equation}
which satisfy the Lie bracket algebra
\begin{equation}\begin{aligned}
\,[L_m, L_n] &= (m-n) \, L_{m+n}, &\qquad  [L_m, T^{(a)}_n] &= - n \, T^{(a)}_{m+n}, \\
[T^{(0)}_m, T^{(\pm)}_n] &= \mp T^{(\pm)}_{m+n}, & [T^{(+)}_m, T^{(-)}_n] &= 2 \, T^{(0)}_{m+n}\, .
\end{aligned}\end{equation}
Thus, the classical asymptotic symmetry algebra is a Witt algebra and an $sl(2,{\mathbb R})$ current algebra.

We use the Brandt--Barnich--Comp\`{e}re (BBC) formulation \cite{Barnich:2001jy, Barnich:2007bf} for
computing the corresponding charges of our geometry. We find that the charges are integrable over
the solution space if $\delta N = 0$ with
\newpage
\bea
\small
\delta \!\!\!/ Q_\xi &=& \frac{1}{8 \pi G} \delta \int d\phi \, \Big\{ B(x^+) \Big[  \kappa (x^+) +  N^2 (\frac{1}{2} f^2(x^+)- g(x^+) \bar g(x^+) ) \cr
&&~~~~~~~~~~~~~~~~~~\,~~~~~~~~~~~~~~~~~~  +\frac{N^2}{2} (e^{i N x^-} g(x^+) + e^{-i N x^-} \bar g(x^+)) \cr
&&~~~~~~~~~~~~~~~~~~\,~~~~~~~~~~~~~~~~~~  + \frac{i}{2} N  \, \partial_+[B (x^+)\, (e^{i N x^-} g(x^+) - e^{-i N x^-} \bar g(x^+))]\Big] \Big\} \cr
&& -\frac{1}{8\pi G} \delta \int d\phi \, N^2 \Big[ \frac{1}{2} A_0 (x^+) f(x^+) - (g(x^+) A(x^+) + \bar g(x^+) \bar A(x^+)) \Big]\, . \nonumber \\
\eea
These can be integrated between the configurations trivially in the solution space from $f(x^+) = g(x^+) = \kappa(x^+) = 0$ to general values of these fields to write down the charges
\bea
\label{aps-charges}
Q_B &=& \frac{1}{8 \pi G}  \int_0^{2\pi} d\phi \, \Big[  B(x^+) \Big( \kappa (x^+) + \frac{N^2}{2}(f^2(x^+)- 2 \, g(x^+)\bar g(x^+)) \Big) \cr
&& ~~~~~~~~~~~~~~~~~~~~~~~~~\,~~~~~~~~~~~~ +\frac{1}{2}(\partial_+ - \partial_-)\partial_- [e^{i N x^-} g(x^+) + e^{-i N x^-} \bar g(x^+)]\Big]  \cr
&=& \frac{1}{8 \pi G}  \int_0^{2\pi} d\phi \, \Big[  B(x^+)[ \kappa (x^+) + \frac{N^2}{2}(f^2(x^+) - 2 \, g(x^+)\bar g(x^+))]  \cr
&& ~~~~~~~~~~~~~~~~~~~~~\,~~~~~~~~~~~~~~~ +\frac{1}{32 \pi G}   \partial_- [e^{i N x^-} g(x^+) + e^{-i N x^-} \bar g(x^+)]\Big|^{\phi = 2\pi}_{\phi = 0} \, ,\nonumber \\
\eea
\bea
Q_A &=& -\frac{N^2}{8\pi G}  \int_0^{2\pi} d\phi \,  \Big[ \frac{1}{2} A_0 (x^+) f(x^+) - (g(x^+) A(x^+) + \bar g(x^+) \bar A(x^+)) \Big] \, .
\eea
The boundary term in \eqref{aps-charges} vanishes as we assumed $g(x^+)$ to be periodic and $N$ to be an integer. The algebra of these charges admits central charges. We find that the central term in the commutation relation between charges corresponding to two asymptotic symmetry vectors $\xi$ and $\tilde \xi$ is given by
\begin{multline}
(-i) \frac{l}{32 \pi G} \int_0^{2\pi} d\phi \, \Big[ B'(x^+) \tilde B''(x^+) - B(x^+) \tilde B'''(x^+) \\ + 2 N^2 \, A_0 (x^+) \tilde A'_0 (x^+) - 4 N^2 \Big(A (x^+) \bar{\tilde A}'(x^+) + \bar A(x^+) \tilde A'(x^+) \Big)\Big].
\end{multline}
These give rise to the following algebra for the charges\footnote{The bracket in \eqref{aps-algebra} is $i$ times the Dirac bracket.}
\bea
\label{aps-algebra}
[L_m, L_n] &=& (m-n) \, L_{m+n} + \frac{c}{12} m^3 \, \delta_{m+n, 0} \,  , \cr
[L_m, T^a_n] &=& - n \, T^a_{m+n} \, , \cr
[T^a_m, T^b_n] &=& {f^{ab}}_c T^c_{m+n} + \frac{k}{2} \eta^{ab} \, m \, \delta_{m+n, 0}
\eea
with 
\bea
c= \frac{3l}{2G}, ~~ k = \frac{c}{6}, ~~ {f^{0+}}_+ = -1,~~ {f^{0-}}_- = 1, ~~ {f^{+-}}_0 = 2, ~~ \eta^{00} = -1, ~~ \eta^{+-} = 2. \nonumber\\
\eea
 This is precisely the $sl(2, {\mathbb R})$ current algebra found in \cite{Polyakov:1987zb}.

\section{Conclusion}

In this note we have provided boundary conditions for 3-dimensional gravity with negative cosmological constant such that the algebra of asymptotic symmetries is an $sl(2, {\mathbb R})$ current algebra. In the process we showed that the boundary term proposed by CSS \cite{Compere:2013bya} admits a more general set of boundary conditions, which enables our result. 

It should be noted that our asymptotic symmetry algebra does contain the full isometry algebra of the global $AdS_3$ solution. This feature is similar to Brown--Henneaux \cite{Brown:1986nw} though one does not demand that the asymptotic vector fields of interest be asymptotically Killing; instead one uses the more general notion of asymptotic symmetries advocated by BBC \cite{Barnich:2001jy, Barnich:2007bf}. Using the BBC formulation, we computed the algebra of charges and found the level $k$ to be $c/6$, independent of the parameter $N$. 

To understand the relation to 2-dimensional induced gravity of Polyakov in light-cone gauge \cite{Polyakov:1987zb} further, it will be interesting to see if the correlation functions of the boundary currents, and the effective action for the dynamical fields of the boundary can also be recovered from the gravity side. See \cite{Banados:2002ey} for a discussion on the latter issue. Of course, connections between 3-dimensional gravity with negative cosmological constant and Liouville theory, which arises as a different gauge-fixing of~\eqref{polyeq1}, are well-known (see e.g.~\cite{Carlip:2005zn}, and the recently proposed boundary conditions in~\cite{Troessaert:2013fma}).

It will be interesting to see how adding matter to $AdS_3$ gravity would generalize our analysis. The boundary conditions of \cite{Compere:2013bya} have been found to be related to string theory solutions of \cite{Azeyanagi:2012zd} with a warped $AdS_3$ factor. It will be interesting to explore whether the boundary conditions in \eqref{aps} also play a role in some string theory context.

The non-linear solution in (\ref{nlsol1}, \ref{nlsol2}) does not contain the conventional positive mass BTZ \cite{Banados:1992wn} black hole. The special case of vanishing charges is given by $f(x^+) = g(x^+) = \bar g(x^+) = \kappa(x^+) = 0$ which is simply an extremal BTZ but with negative mass (in global $AdS_3$ vacuum). The comments of CSS \cite{Compere:2013bya} about the possible existence of ergoregions and instabilities in their solution also apply to \eqref{nlsol1}. It will be important to understand these issues better.

Finally, it is intriguing that different ways of gauge-fixing the induced gravity \cite{Polyakov:1987zb} lead to different boundary conditions in the bulk and therefore apparently different holographic duals. It will be important to understand the class of physical theories one can obtain this way and how they are related to each other.

%\newpage

\bibliographystyle{utphys}
%\bibliography{aps01}
\providecommand{\href}[2]{#2}\begingroup\raggedright\endgroup

\end{document}